\documentstyle[12pt]{article}
\newcommand{\ce}{{{\cal E}}}

\begin{document}

\title{ Concept of semi-velocity  and related wave equation }

\author{
Robert M. Yamaleev\\
Joint Institute for Nuclear Research,\\
Laboratory of Informational Technology,\\
Dubna, Russia.\\
 Email: yamaleev@jinr.ru }
 \maketitle
\begin{abstract}

The semi-velocity is defined as an exponential function of the
rapidity. Physically the semi-velocity is interpreted as the
relativistic analogue of the phase velocity, a geometrical
interpretation done within the framework of Beltrami-Klein and
Poincar\'e models. A kinematical wave equation related with the
Fermat principle and the concept of the semi-velocity is  derived.

\end{abstract}

\section{Introduction}

The relativistic kinematics is a part of the relativistic
mechanics independent of the inertial mass, $m$. The main notion
of this theory, evidently, is the velocity, $V$. In the
relativistic mechanics the dynamical notions like energy $cp_0$
and momentum $p$ are definite functions of the mass and the
velocity and, usually, they are proportional to the proper mass
$m$.  The principal relationship of the relativistic mechanics is
the mass-shell equation
$$
p_0^2-p^2=m^2c^2. \eqno(1.1)
$$
Factorize this equation with respect to the square of the momentum
$$
p^2=(p_0-mc)(p_0+mc).  \eqno(1.2)
$$
In the kinematics instead of the product of the quantities
 the quotients of these quantities are used, e.g.,
the role of the momentum $p^2$ defined by the product of the
quantities $(p_0\pm mc)$, in the kinematics we define the analogy
of the square of the momentum as the following quotient
$$
k^2=\frac{p_0-mc}{p_0+mc}. \eqno(1.3)
$$
In the relativistic kinematics the velocity is defined by the
quotient of the energy-momentum
$$
V=c\frac{p}{p_0}.
$$
Parametrization of the energy-momentum by the velocity leads us to
the formulae proportional to the proper mass:
$$
p_0=\frac{mc}{\sqrt{1-\frac{V^2}{c^2}}},~
p=\frac{mV}{\sqrt{1-\frac{V^2}{c^2}}}.  \eqno(1.4)
$$
In terms of the velocity  the quotient in (1.3) is written as
$$
k^2=\frac{1-\sqrt{1-\frac{V^2}{c^2}}}{1+\sqrt{1-\frac{V^2}{c^2}}}.
\eqno(1.5)
$$
Evidently, this formula is the squared form of the equation
$$
k=\frac{c^2}{V}(1\pm \sqrt{1-\frac{V^2}{c^2}}). \eqno(1.6)
$$

In the period of construction a bridge between wave mechanics and
the classical mechanics the concept {\it analogue of the phase
velocity } in the classical mechanics  had played a central role.

 On the basis of Hamilton-Jacobi equations E.Schrodinger
has defined this quantity as follows \cite{Schrodinger}
$$
v=\frac{\ce}{p}, \eqno(1.7)
$$
where $p$ is the momentum  and $\ce$ is the total energy of a
particle with mass $m$,
$$
\ce=\frac{p^2}{2m}+U(r). \eqno(1.8)
$$
Evidently, the analogue of the phase velocity differs of
definition of the velocity in the classical mechanics. In absence
of the external fields the phase velocity $v$ differs of the
velocity $V$ by the factor one-half:
$$
v=\frac{p^2/2m}{p}=\frac{1}{2}V.  \eqno(1.9)
$$

By the de Broglie's hypothesis  the phase velocity is defined as
\cite{Broglie}
$$
v_{phase}=\frac{c^2}{V}, \eqno(1.10)
$$
where $V$ is the velocity of the particle regardless of wave
behavior. In the case of free motion the  formula of de Broglie
(1.1) can be considered as a relativistic generalization of the
Schrodinger definition (1.7). In fact, if $cp_0$ is relativistic
expression for the kinetic energy, then,
$$
c\frac{p_0}{p}=\frac{c^2}{V}. \eqno(1.11)
$$

In the relativistic kinematics the velocity admits a
trigonometrical representation via an hyperbolic angle
$$
V=c~\tanh(\psi).  \eqno(1.12)
$$
In this representation  the  phase velocity  corresponds one half
of the hyperbolic angle
$$
v=c~\tanh(\frac{1}{2}\psi). \eqno(1.13)
$$
Furthermore, denoting the relativistic summation formula for
velocities by the symbol
$$
V=(v_1\bigoplus v_2),  \eqno(1.14)
$$
the formula (1.4) will be re-written as
$$
V=(v\bigoplus v)=\frac{v+v}{1+v^2/c^2}=\frac{2v}{1+v^2/c^2}.
\eqno(1.15)
$$
This formula  we take as a basic formula of the wave kinematics.
In the relativistic dynamics we propose to denominate this
velocity as {\it semi-velocity}. The concept of the semi-velocity
has its remarkable interpretation via the cross-ratio
\cite{Yamaleev4} and distance between two points in the hyperbolic
space. This concept leads to new kind of wave equation which can
be interpreted as {\it wave equation of the relativistic
kinematics}.

The paper is presented by the following sections.

Section 2 we introduce a concept of the semi-velocity and give its
geometrical interpretation. In Section 3 define an analogue of the
phase velocity in the relativistic mechanics and suggest a new
kind of wave equation for the massive particle.

\section{  Geometrical interpretations of the semi-velocity}

Let $V$ be a velocity of the point-particle defined with respect
to coordinate time expressed via the hyperbolic angle $\psi$, the
rapidity,
$$
V=c\tanh(\psi). \eqno(2.1)
$$
Define the velocity $v$ at half-rapidity $\psi/2$ by the same
trigonometric function
$$
v=c\tanh(\psi/2). \eqno(2.2)
$$
Then, the velocity $V$ is the function of $(v)$ of the form
$$
V=(v\bigoplus v)=\frac{2v}{1+v^2/c^2}. \eqno(2.4)
$$
The quantity $v$ we will denominate as the {\it semi-velocity}. In
general, we introduce two types of the semi-velocities by
$$
v_-v_+=c^2,~~v_-=v. \eqno(2.5)
$$
Formula (2.4) is valid for the both velocities:
$$
V=(v_{\pm}\bigoplus v_{\pm})=\frac{2v_{\pm}}{1+v_{\pm}^2/c^2}.
\eqno(2.6)
$$
Equation (2.6) is converted into a quadratic equation with respect
to $v_{\pm}$:
$$
v^2-\frac{2c^2}{V}v+c^2=0,  \eqno(2.7)
$$
with roots
$$
\frac{v_{\pm}}{c}=\frac{c}{V}(1\pm \sqrt{1-\frac{V^2}{c^2}}).
\eqno(2.8)
$$
In terms of the dynamical variables the semi-velocity is defined
by the formulae
$$
\frac{v_{\pm}}{c}=\frac{p}{p_0\mp mc}=\frac{p_0\pm mc}{p},
\eqno(2.9)
$$
correspondingly, the square of this equation is
$$
\frac{v_{\pm}^2}{c^2}=\frac{p_0 {\pm}mc}{p_0\mp mc}.  \eqno(2.10)
$$
In addition, let us observe a connection of the semi-velocities
with the {\it Pythagoras spinors} $(q_1, q_2)$. The trio of
dynamic variables, $p_0,p,mc$ are bilinear functions of the
Pythagoras spinors
$$
p^2=q^2_1q^2_2,~p_0=\frac{1}{2}(q^2_1+q^2_2),~~mc=\frac{1}{2}(q^2_1-q^2_2).
\eqno(2.11)
$$
  The semi-velocities are defined simply as fractions of the
Pythagoras spinors
$$
\frac{v^2_+}{c^2}=\frac{q^2_1}{q^2_2},~~\frac{v^2_-}{c^2}=\frac{q^2_2}{q^2_1}.
\eqno(2.12)
$$

It is remarkable that the formula for semi-velocity admits
exponential representation because \cite{Yamaleev3}
$$
\exp(2mc\phi)=\frac{p_0+mc}{p_0-mc}. \eqno(2.13)
$$
Then,
$$
\exp(mc\phi)=\coth(\psi/2)=v_+,~~v_-=\exp(-mc\phi). \eqno(2.14)
$$

In order to give a geometrical interpretation of the semi-velocity
let us recall some elements of the Beltrami-Klein and the
Poincar\'e models of the hyperbolic geometry.

The Klein disk model (also known as the Beltrami–Klein model) and
the Poincar\'e disk model are both models that project the whole
hyperbolic plane in a disk. The two models are related through a
projection on or from the hemisphere model. The Klein disk model
is an orthographic projection to the hemisphere model while the
Poincar\'e disk model is a stereographic projection.

When projecting the same lines in both models on one disk both
lines go through the same two ideal points (the ideal points
remain on the same spot), also the pole of the chord in the Klein
disk model is the center of the circle that contains the arc in
the Poincar\'e disk model.

A point $(x,y)$ in the Poincar\'e disk model maps to
$$ \left({\frac
{2x}{1+x^{2}+y^{2}}},~~ {\frac {2y}{1+x^{2}+y^{2}}}\right)
\eqno(2.15)
$$
in the Klein model.

A point (x,y) in the Klein model maps to
$$
\left({\frac {x}{1+{\sqrt {1-x^{2}-y^{2}}}}}\ ,\ \ {\frac
{y}{1+{\sqrt {1-x^{2}-y^{2}}}}}\right) \eqno(2.16)
$$
in the Poincar\'e disk model.

If ${\displaystyle u}$  is a vector of norm less than one
representing a point of the Poincar\'e disk model, then the
corresponding point of the Klein disk model is given by
$$
{\displaystyle s={\frac {2u}{1+u\cdot u}}.}  \eqno(2.17)
$$
Conversely, from a vector $s$  of norm less than one representing
a point of the Beltrami–Klein model, the corresponding point of
the Poincar\'e disk model is given by:
$$
u={\frac {s}{1+{\sqrt {1-s\cdot s}}}}={\frac {\left(1-{\sqrt
{1-s\cdot s}}\right)s}{s\cdot s}}. \eqno(2.18)
$$

Thus, if $P$ is a point a distance $V/c$ from the center of the
unit circle in the Beltrami-Klein model, then corresponding point
in the Poincar\'e disk model a distance of $v/c$ of the same
radius:
$$
\frac{v}{c}=\frac{c}{V}(1- \sqrt{1-\frac{V^2}{c^2}}). \eqno(2.19)
$$
Conversely, if $P$ is a point a distance $v/c$ from the center of
the unit circle in the Poincar\'e disk model, then the
corresponding point of the Beltrami–Klein model is a distance of
$V/c$ on the same radius:
$$
\frac{V}{c}=\frac{2v/c}{1+v^2/c^2} . \eqno(2.20)
$$
 By taking into account (2.9) we get
$$
\frac{v_-}{v_+}=\exp(2\psi)=\frac{1-\sqrt{1-V^2/c^2}}{1+\sqrt{1-V^2/c^2}}.
\eqno(2.21)
$$
This formula  coincides with the formula for distance between two
points on the hyperbolic plane (see, for instance, \cite{Kagan},
p.142). In this geometry distance between two point is defined by
logarithmic formula
$$
\rho=\frac{k}{2}\log\frac{1+\sqrt{1-J^2}}{1-\sqrt{1-J^2}},
\eqno(2.22)
$$
where $J$ in coordinates Beltrami is defined by the formula
$$
J^2=\frac{(z_1^2-x_1^2-y_1^2)(z^2_2-x^2_2-y_2^2)}{(z_1z_2-x_1x_2-y_1y_2)^2}.
\eqno(2.23)
$$
Formula (2.22) written in the exponential form coincides with
(2.20), hence
$$
V^2/c^2=J^2. \eqno(2.24)
$$

If the equation of the fundamental conic is
$$
\Omega=\sum^3_{i,j} a_{ij}x_iy_j=0,  \eqno(2.25)
$$
then the distance between the two points $x$ and $y$, in
homogeneous coordinates $(x_1,x_2,x_3)$ and $(y_1,y_2,y_3)$
$$
\rho=\frac{k}{2}\log\frac{\Omega_{xy}+\sqrt{\Omega_{xy}^2-\Omega_{xx}\Omega_{yy}}}{\Omega_{xy}-\sqrt{\Omega_{xy}^2-\Omega_{xx}\Omega_{yy}}},
\eqno(2.26)
$$
Here,
$$
\Omega_{kj}=z_kz_j-x_kx_j-y_ky_j  \eqno(2.27)
$$
then, the following identity holds true
$$
\Omega_{11}\Omega_{22}-{\Omega^2_{12}}=(x_1y_2-x_2y_1)^2-(y_1z_2-y_2z_1)^2-(x_1z_2-x_2z_1)^2.
\eqno(2.28)
$$
In this model, the velocity $V$ is presented by formula
$$
V^2/c^2=\frac{\Omega_{11}\Omega_{22}}{\Omega^2_{12}}, \eqno(2.29)
$$
and the rapidity $\xi$ with the distance $\rho$ and speed of light
with curvature $k$. According to identity (2.28), the energy
$p_0$, momentum $p$ and the mass $mc$ can be identified as follows
$$
p^2=(z_1^2-x_1^2-y_1^2)(z^2_2-x^2_2-y_2^2),~~p_0=z_1z_2-x_1x_2-y_1y_2,
$$
$$
~~m^2c^2=(x_1y_2-x_2y_1)^2-(y_1z_2-y_2z_1)^2-(x_1z_2-x_2z_1)^2..
\eqno(2.30)
$$

{\bf 2.2 Semi-velocity and Chebyshev polynomials. }

The Chebyshev polynomials of the first kind are defined by  the
recurrence relations \cite{Chebysh1}
$$
T_0(x)=1,~~T_1(x)=x,~~T_{n+1}(x)=2xT_n(x)-T_{n-1}(x).
$$
The corresponding difference equation can be associated with unit
defined by the equation \cite{Dattoli1}
$$
H^2=2xH-1,~~H_{\pm}=x\pm \sqrt{x^2-1}, \eqno(2.31)
$$
and
$$
H^n_{\pm}=A_n+H_{\pm}B_n,  \eqno(2.32)
$$
where $A_n$ and $B_n$ are functions of the variable $x$ satisfying
recurrences that can be presented in the matrix form
$$
\left( \begin{array}{c}
A_{n+1}\\
B_{n+1}
\end{array} \right)=
\left( \begin{array}{cc}
0&-1\\
1&2x
\end{array} \right)
\left( \begin{array}{c}
A_n\\
B_n
\end{array} \right). \eqno(2.33)
$$
Denote
$$
\frac{c}{V}=x. \eqno(2.34)
$$
Then the formula (2.8) for semi-velocity coincides with $H_{\pm}$:
$$
\frac{v_{\pm}}{c}=(x\pm \sqrt{x^2-1} )=H_{\pm}. \eqno(2.35)
$$
This fact prompts us to construct the Chebyshev sequences from the
components of the semi-velocity. The Chebyshev polynomials of
first kind are given by
$$
T_n(x)=\frac{1}{2c}(v^n_++v^n_-).  \eqno(2.36)
$$
The semi-velocity as the function of the velocity satisfy the
equation
$$
\frac{d}{dx}v_{\pm}=\pm\frac{v_{\pm}}{\sqrt{x^2-1}},  \eqno(2.37)
$$
The Chebyshev polynomials of the first kind in the study of
differential equations arises as solutions of the equation
$$
[(1-x^2)\partial_x^2-x\partial_x+n^2]T_n(x)=0.  \eqno(2.38)
$$
And the Chebyshev equations of the second kind
$$
U_n(x)=\frac{H_{+}^{n+1}-H_{-}^{n+1}}{2\sqrt{x^2-1}}  \eqno(2.39)
$$
verify the following differential equation
$$
[(1-x^2)\partial_x^2-3x\partial_x+n(n+2)]U_n(x)=0.  \eqno(2.40)
$$
The exponential generating function is
$$
\sum_{n=0}^{\infty}T_n(x)\frac{t^n}{n!}=\frac{1}{2}(e^{v_-t/c}+e^{v_+t/c}).
\eqno(2.41)
$$

\section{ Derivation of the wave equation from Fermat's principle }

{\bf The Fermat law.}

The wave equation of the quantum mechanics, the Schrodinger
equation, had been obtained on the basis of Hamilton principle and
the Hamilton-Jacobi equation \cite{Persico}. Our goal is to pass
the same way, but by using as an starting platform the Fermat
principle. In refs. \cite{Oziewicz1} and \cite{Adan} it has been
established that for the electromagnetic waves radiating in a
medium the velocities $v_{\pm}$ are related with refractive index
$n$ according to the following formulae
$$
\frac{v_-^2}{c^2}=\frac{p_0-mc}{p_0+mc}=\frac{1}{n^2},
\eqno(3.1,a)
$$
for $n>1$, and
$$
\frac{v_+^2}{c^2}=\frac{p_0+mc}{p_0-mc}=\frac{1}{n^2},
\eqno(3.1,b)
$$
for $n<1$.

Consequently, the velocities $v_{\pm}$ have to be interpreted as
the phase velocities related in ordinary way with refractive index
$n$.

As a first step let us recall as the definition of analogue of the
phase velocity had been used in the procedure of modification of
the Eikonal equation into the Hamilton -Jacobi equation. Let us
start with the Eikonal equation in a geometrical wave theory of
the form
$$
\frac{d W}{dl}=n=\frac{c}{v}, \eqno(3.2)
$$
admits direct integration.  By taking into account
$dl/dt=v,~dl/v=dt$, integrate equation (3.2) to obtain
$$
W/c=\int^2_1 \frac{dl}{v}= \int^2_1 dt=t_2-t_1.
$$
Thus, the value $W/c$ is an interval of time $T_{21}=t_2-t_1$
during of which the wave front passes given distance
$d_{12}=l_1-l_2$.

The time of radiation of the light in the medium with refractive
index $n$ is given by the integral
$$
T=\int \frac{dl}{v(x,y,z)}=\int \frac{n}{c}dl.  \eqno(3.3)
$$
According to Fermat's law this time has to be minimal (extremal):
$$
\delta T=0.  \eqno(3.4)
$$

Now, instead of the phase velocity $v$ let us use its expression
via the velocity $V$ defined with respect to the coordinate time.
The function $v=v(V)$ is given in the explicit form by
$$
\frac{v}{c}|_{\pm}=\frac{c}{V}(1\pm \sqrt{1-\frac{V^2}{c^2}}).
\eqno(3.5)
$$
Substitute this formula into the integral
$$
W=\int^2_1 \frac{Vdl}{c(1+ \sqrt{1-\frac{V^2}{c^2}})}=\int^2_1
\frac{V^2dt}{c(1+ \sqrt{1-\frac{V^2}{c^2}})}=\int^2_1 dt~((1-
\sqrt{1-\frac{V^2}{c^2}})).  \eqno(3.6)
$$
 Inside  the integral we have the Lagrange function
$$
 L=1- \sqrt{1-\frac{V^2}{c^2}},  \eqno(3.7)
$$
which differs of the relativistic Lagrangian only by the constant.

{\bf 3.2 The Hamilton-Jacobi equation. }

With aim to adopt the eikonal equation (3.2) to the dynamics of
the particle, first of all, let us use relationship of refractive
index of the medium express with the phase velocity $v$,
$$
{n^2}=\frac{c^2}{v^2}. \eqno(3.8)
$$
Perform  replacement on making use of formula (1.1) in (2.1), this
gives,
$$
(\nabla W)^2=\frac{c^2}{v^2}\rightarrow  (\nabla
\frac{W}{c})^2=\frac{p^2}{\ce^2} \rightarrow (\nabla (\ce
W/c))^2=p^2. \eqno(3.9)
$$
For the simplest case of the inertial free motion define
$$
W=-v_0t+r\frac{v_0}{v}, \eqno(3.10)
$$
so that,
$$
\frac{\partial W}{\partial t}=-v_0,~~\frac{\partial W}{\partial
r}=\frac{v_0}{v}. \eqno(3.11)
$$
Define the function of action by
$$
S=\ce W/c=-\ce t+pr.  \eqno(3.12)
$$
Differential form of this equation
$$
dS=-\ce dt+p dr,  \eqno(3.13)
$$
offers formulas
$$
\frac{\partial S}{\partial t}=-\ce,~~\frac{\partial S}{\partial
r}=p.  \eqno(3.14)
$$
Substituting these definitions into formula for the energy (1.2)
we come to Hamilton-Jacobi equations
$$
|\nabla S|^2=2m(\ce-U),~~\frac{\partial S}{\partial t}=-\ce,
\eqno(3.15)
$$
From Hamilton-Jacobi theory it follows that the wave front of the
function of action $S$ is moving with the phase velocity
$$
v=\frac{\ce}{\sqrt{2m(\ce-U)}}, \eqno(3.16)
$$
in fact,
$$
\frac{dS}{dt}=\frac{\partial S}{\partial
t}+\frac{dr}{dt}~\frac{\partial S}{\partial r}=0 \rightarrow
\frac{dr}{dt}=-\frac{\partial S}{\partial t}~/~\frac{\partial
S}{\partial r}=\frac{\ce}{p}. \eqno(3.17)
$$

{\bf 3.3 The kinematic Wave equation. }

Transformation of the Eikonal equation into relativistic
Hamilton-Jacobi equation is worked out in a similar way. Write
$$
(\nabla W)^2=\frac{c^2}{v^2}=\frac{p^2}{\ce^2},  \eqno(3.18)
$$
where
$$
\ce=cp_0+U(r), \eqno(3.19)
$$
and
$$
p^2=(\ce-U(r)-mc^2)(\ce-U(r)+mc^2).  \eqno(3.20)
$$
where
$$
\frac{\partial S}{\partial t }=-\ce.~ \eqno(3.21)
$$
Since
$$
n^2=\frac{v^2}{c^2}=\frac{{p_0-mc}}{{p_0+mc}}=
\frac{{\ce/c+mc-U(r)/c}}{{\ce/c-mc-U(r)/c}},  \eqno(3.22)
$$
 The Eikonal equation we present of the form
$$
(\nabla W)^2= n^2=
\frac{p_0-mc}{p_0+mc}=\frac{{\ce/c+mc-U(r)/c}}{{\ce/c-mc-U(r)/c}}.
\eqno(3.23)
$$
This equation can be understood as an analogue of the relativistic
Hamilton-Jacobi equation
$$
(\nabla S)^2=
p^2=({p_0-mc})({p_0+mc})=(\ce/c+mc-U(r)/c)(\ce/c-mc-U(r)/c).
\eqno(3.24)
$$

The wave equation in the medium is written as
$$
(\nabla^2+n^2\kappa_0^2)\Psi=0,  \eqno(3.25)
$$
$$
\kappa_0=\frac{2\pi}{\lambda_0}. \eqno(3.26)
$$
Substitute here the definition of $n^2$ via the semi-velocity
$k^2$ from (1.5). We get
$$
\nabla^2\Psi+\kappa_0^2
\frac{1-\sqrt{1-\frac{V^2}{c^2}}}{1+\sqrt{1-\frac{V^2}{c^2}}}
\Psi=0. \eqno(3.27)
$$
In presence of the external potential field $U(r)$ on making use
of the formula $\ce=p_0c+U(r)$ we arrive at the following wave
equation
$$
\nabla^2\Psi+\kappa_0^2\frac{\ce-mc^2-U(r)}{\ce+mc^2-U(r)}\Psi=0.
\eqno(3.28)
$$
In accordance with (3.15) we write
$$
(\frac{\partial W}{\partial r})^2=(\frac{c}{v})^2 (\frac{\partial
W}{\partial t})^2  . \eqno(3.29)
$$
Then, the corresponding wave equation is formulated as
$$
\nabla^2\Psi+\frac{\ce-mc^2-U(r)}{\ce+mc^2-U(r)}\frac{\partial^2
}{c^2\partial t^2} \Psi=0. \eqno(3.30)
$$

\section{ Concluding remarks}

The formula of the semi-velocity had been used in various branches
of the relativistic dynamics and  models of the hyperbolic
geometry. Furthermore, in these models the notion of the
semi-velocity had played a central role \ cite{hand}.  Physics of
the accelerators, colliders,  is one of the examples where the
concept of the semi-velocity as the relative velocity of
particles, it has been introduced. We have shown that the
semi-velocity is related:

(1) with the refractive index of the medium;

(2) with the velocity of radiating electro-magnetic  field;

(3) with the phase velocity of the massive elementary particle;

(4) with the cross-ratio;

(5) with the distance in the hyperbolic geometry.

Connection "phase velocity---semi-velocity" leads to a new kind of
the wave equation, connection "semi-velocity--- cross-ratio" leads
to geometrical interpretation of the relativistic kinematics.

\end{document}